\documentclass[preprint,authoryear,12pt]{elsarticle}

\usepackage{url}
\usepackage{subfigure}
\usepackage{graphicx}
\usepackage{amsmath,amsfonts,latexsym,amssymb,euscript}
\usepackage{algorithm}
\usepackage{algorithmic}

\newcommand{\bx}{\mbox{\boldmath $x$}}

\newcommand{\bA}{\mbox{\boldmath $A$}}
\newcommand{\bB}{\mbox{\boldmath $B$}}

\newcommand{\bD}{\mbox{\boldmath $D$}}

\newcommand{\bX}{\mbox{\boldmath $X$}}

\newcommand{\bl}{\mbox{\boldmath $l$}}
\newcommand{\bL}{\mbox{\boldmath $L$}}

\newcommand{\br}{\mbox{\boldmath $r$}}
\newcommand{\bR}{\mbox{\boldmath $R$}}

\newcommand{\bb}{\mbox{\boldmath $b$}}

\newcommand{\bz}{\mbox{\boldmath $z$}}

\newcommand{\hbX}{\hat{\mbox{\boldmath $X$}}}
\newcommand{\hbA}{\hat{\mbox{\boldmath $A$}}}
\newcommand{\hbB}{\hat{\mbox{\boldmath $B$}}}

\journal{Statistics and Probability Letters}
\begin{document}
\begin{frontmatter}

\title{A Very Fast Algorithm for Matrix Factorization}

\author[label1]{Vladimir Nikulin}

\author[label2]{Tian-Hsiang Huang}

\author[label3]{Shu-Kay Ng}

\author[label1,label4]{Suren I Rathnayake}

\author[label1,label4]{Geoffrey J McLachlan\corref{cor1}}
\ead{gjm@maths.uq.edu.au}

\address[label1]{Department of Mathematics,
University of Queensland,
St Lucia, QLD, Australia}

\address[label2]{Institute of Information Management,
National Cheng Kung University,
Tainan, Taiwan}

\address[label3]{School of Medicine,
Griffith University,
Meadowbrook, QLD, Australia}

\address[label4]{Institute for Molecular Bioscience, 
University of Queensland,
St Lucia, QLD, Australia}

\cortext[cor1]{Corresponding author}

\begin{abstract}
We present a very fast algorithm for general matrix factorization
of a data matrix 
for use in the statistical analysis of 
high-dimensional data via latent factors. 
Such data are prevalent across many application areas and
generate an ever-increasing demand for methods of dimension
reduction in order to undertake the statistical analysis of interest.
Our algorithm uses a gradient-based approach which can be used with an
arbitrary loss function provided the latter is differentiable.
The speed and effectiveness of our algorithm for dimension reduction
is demonstrated in the context of supervised classification of some real
high-dimensional data sets from the bioinformatics literature.
\end{abstract}


\end{frontmatter}


\section{Introduction}

We let $\bx_1,\,\dots,\,\bx_n$ denote $n$ observed $p$-dimensional
observations, where the number of variables $p$ is very large relative to
$n$. For example, in the analysis of microarray gene-expression data,
$n$ (the number of tissues) might be only 50, whereas $p$ (the number of 
genes) might be in the tens of thousands.
We follow the traditional biologists' practice of letting
$$\bX=(\bx_1,...,\bx_n)$$
be the $p \times n$ data matrix.
The usual statistical practice is to take the transpose of $\bX$, $\bX^T$,
as the data matrix.
Without loss of generality, we assume that the overall mean of $\bX$ is
zero.

In most statistical analyses of the data matrix $\bX$, some form of dimension
reduction is required, typically before the primary analysis is performed,
or with some approaches it might be done in conjunction with the main
analysis.
In recent times, much attention has
been given to matrix factorizations of the form,
\begin{equation}
\bX=\bA \bB,
\label{eq:1}
\end{equation}
where $\bA$ is a $p \times q$ matrix and $\bB$ is a $q \times n$ matrix
and where $q$ is chosen to be much smaller than $p$. For a specified value
of $q$, the matrices $\bA$ and $\bB$ are chosen to minimize 
\begin{equation}
\|\bX-\bA\bB\|^2,
\label{eq:2}
\end{equation}
where $\|\cdot\|$ is the 
Frobenius norm (the sum of squared elements of the matrix).  
With this factorization, dimension reduction is effected by replacing the
data matrix $\bX$ by the solution $\hbB$ for the factor matrix $\bB$; the
$i$th row of $\hbB$ gives the values of the $i$th metavariable
for the $n$ entities.
Thus the original $p$ variables are replaced by $q$ metavariables.
When the elements 
of $\bX$ are nonnegative, we can restrict the elements of $\bA$ and $\bB$
to be nonnegative.  This approach is called nonnegative matrix
factorization (NMF) in the literature (Lee and Seung, 1999). 
We shall call the general approach 
where there are no constraints on $\bA$ and $\bB$,  GMF (general matrix 
factorization).  

The classic method for factoring the data matrix $\bX$ is singular-value
decomposition (SVD, Golub and van Loan (1983)).
It follows from this theorem that we can decompose $\bX$ exactly into the form
\begin{equation}
\bX=\bL \bD \bR^T,
\label{eq:3}
\end{equation}
where $\bL =(\bl_1,\ldots,\,\bl_k)$ 
is a $p \times k$ matrix with orthonormal columns, 
$\bR=(\br_1,\,\ldots,\,\br_k)$ is a
$n \times k$ matrix with orthonormal columns, 
$\bD$ is a diagonal matrix with
elements $d_1\geq d_2\geq \cdots \geq d_k >0$, and 
$k \leq \min (p, n)$ is the rank of $\bX$.
For any $q\leq k$, 
\begin{equation}
\sum_{i=1}^q d_i \bl_i \br_i^T =\arg \min_{\hbX \in M(q)} \|\bX-\hbX\|^2,
\label{eq:4}
\end{equation}
where $M(q)$ is the set of rank-$q$ $p \times n$ matrices; see, for
example, Eckart and Young (1936).

Let $\bL^{(q)}=(\bl_1,\,\ldots,\,\bl_q)$,
$\bR^{(q)}=(\br_1,\,\ldots,\,\br_q)$, and $\bD^{(q)}$
be the diagonal matrix with diagonal elements $d_1,\,\ldots,\,d_q.$
Then on considering the matrix factorization (\ref{eq:1}) of $\bX$, 
it follows from (\ref{eq:4}) that for a specified value of $q$ 
we can find the factor matrices $\bA$ and $\bB$ that minimize
(\ref{eq:2}) by taking
$\hbA=\bL^{(q)}$ and $\bB=\bD^{(q)} \bR^{(q)^T}$.

The calculation of the exact SVD of the matrix $\bX$ has time
complexity
$O(\min\{pn^2,n^2p\})$. 
Hence the use of SVD for high-dimensional data sets is not feasible and
the use of the best $q$-approximation (\ref{eq:4}) for $q$ larger enough
to capture most of the variance in $\bX$ requires essentially the same amount of
time as the full SVD.

Hence we consider a very fast approach to the general matrix factorization
(\ref{eq:1}), using a gradient-based algorithm applicable for an arbitrary
(differentiable) loss function. In the sequel, we consider the exponential
family of loss functions that include the least-squares
loss function (\ref{eq:2}) as a limiting case.
The novelty of our algorithm lies in the way that on each global iteration
it 
\begin{itemize}
\item[(a)] iterates on only a small subset of the elements of the factor matrix
$\bA$ with the other factor matrix $\bB$ fixed before reversing their roles;
\item[(b)] loops through all the terms in the objective function, 
minimizing them individually 
at a time rather than their total sum 
(that is, it adopts a stochastic gradient descent approach).
\end{itemize}

As to be presented in Section 3,  
our algorithm takes only between 10 and 15 seconds 
in performing 300 global iterations
to provide a $q=11$ rank factorization of a 2000 x 62 data matrix
for the colon cancer data set of Alon et al.\ (1999).
In contrast, 20 global iterations with non-negative matrix factorization
(NMF) for the same task required about 25 minutes.

The effectiveness of our algorithm is to be demonstrated in its
application to provide a reduction in the number of genes for use
in the formation of classifiers in the
supervised classification of five well-known high-dimensional data sets 
in the bioinformatics literature.

\section{Background}

Here we consider the factorization of $\bX$ into $\bA\bB$ in the spirit
that it has no real importance in and of
itself other than  as a computationally convenient means for obtaining 
a reduction in  the number of variables.  
Of course in some situations in practice
once the factorization has been made, attention will turn 
to the interpretability of the metavariables.

The latter consideration has led to much recent interest in the use of NMF 
in the analysis of data for which the elements are nonnegative. It constrains
the elements of the factor matrices $\bA$ and $\bB$ to be nonnegative, which
can be advantageous from the point of view of interpretability. 
{Lee and Seung, 1999; Lee and Seung 2001} developed NMF in order to improve upon the
interpretability of the SVD. The nonnegativity constraints on $\bA$ and $\bB$
form a whole in a nonsubtractive way. In this way, NMF is considered as a
procedure for learning a parts-based representation (Lee and Seung, 1999).
However, as pointed out in Li et al.\ (2001)
the additive parts by NMF are not necessarily localized.
This led them to propose a subspace method, called local nonnegative matrix
factorization (LNMF) for learning spatially localized, parts-based representation
of visual patterns; see also (Donoho et al., 2004; Gao et al., 2005; 
Gogel et al., 2007)  and the recent monograph
(Cichocki et al., 2010).

More recently, Ding et al.\ (2010) has considered variations of NMF where the
elements of $\bA$, but not of $\bB$, 
are constrained to be nonegative, and so allowing the data matrix
$\bX$ to have mixed signs (semi-NMF). 
They also consider algorithms in which the basis
vectors of $\bA$ are constrained to be convex combinations of the data points.
In other work, Witten et al.\ (2009) have proposed a penalized matrix
decomposition for computing a $q$-rank approximation to $\bX$. 

\section{Gradient-Based Algorithm for GMF} 

We now describe our gradient-based algorithm for carrying out the general matrix
factorization (\ref{eq:1}) of the data matrix $\bX$.
The objective function to be minimized is given by
\begin{equation} 
L(\bA, \bB) = \frac{1}{p \cdot n} \sum_{i=1}^p \sum_{j=1}^n \Psi(E_{ij}),
\label{eq:G1}
\end{equation}
where $E_{ij} = x_{ij} - \sum_{f=1}^q a_{if} b_{fj}$, and $\Psi$ is the 
loss function assumed to be differentiable with derivative denoted by $\psi$.
For illustrative purposes, 
we take $\Psi$ to be a member of the exponential family of loss functions given by
\begin{equation}
\Psi(x;\, \alpha) = 2 \frac{\left( \cosh(\alpha x) - 1 \right)}{\alpha^2} =
\alpha^{-2} \left( \exp(\alpha x) + \exp(-\alpha x) - 2 \right),
\label{eq:G2}
\end{equation}
where $\alpha$ is a regularization parameter.
Note that a squared loss function may be regarded as a marginal limit in relation to
this family of loss functions since
\begin{equation}
\lim_{\alpha \to 0} \Psi(x;\, \alpha) = x^2.
\label{eq:G3}
\end{equation}
In our initial experiments Nikulin and  McLachlan (2009), we tried  a range of values 
between 0.003 and 0.004 for $\alpha$, 
which gave similar results as for the squared loss function.

The algorithm can be implemented as follows.

\vspace{.2cm}

\renewcommand{\theenumi}{\arabic{enumi}}
\renewcommand{\labelenumi}{\theenumi:}
\noindent
\textbf{ Gradient-based framework for matrix factorization}
\begin{enumerate}
\item Input: $\bX$ - matrix of microarrays.
\item Select $m$ - number of global iterations; $q$ - number of
factors;  
$\lambda > 0$ - initial learning rate, $0 < \xi < 1$ - correction rate, 
$L_S$ - initial value of the target function.
\item Initial matrices $\bA$ and $\bB$ may be generated randomly.
\item Global cycle: repeat $m$ times the following steps 5 - 17:
\item genes-cycle: for $i = 1$ to $p$ repeat steps 6 - 15:
\item tissues-cycle: for $j = 1$ to $n$ repeat steps 7 - 15:
\item compute prediction $S = \sum_{f=1}^q a_{if} b_{fj}$;
\item compute error of prediction: $E = x_{ij} - S$;
\item internal factors-cycle: for $f = 1$ to $q$ repeat steps 10 - 15:
\item compute $\alpha = a_{if} b_{fj}$;
\item update $a_{if} \Leftarrow a_{if} + \lambda \psi(E) b_{fj}$;
\item $E \Leftarrow  E + \alpha - a_{if} b_{fj}$;
\item compute $\alpha = a_{if} b_{fj}$;
\item update $b_{fj} \Leftarrow b_{fj} + \lambda \psi( E) a_{if}$;
\item $E \Leftarrow  E + \alpha - a_{if} b_{fj}$;
\item compute $L = L(\bA, \bB)$;
\item $L_S = L$ if $L < L_S$; otherwise: $\lambda \Leftarrow \lambda \cdot
\xi$.
\item Output: $\bA$ and $\bB$ -- matrices of loadings and metagenes.
\end{enumerate}

The following partial derivatives are necessary 
for the above algorithm (see steps 11 and 14 above):


\begin{eqnarray}
\frac{\partial \Psi(E_{ij})}{\partial a_{if}} &=& -\psi(E_{ij}) b_{fj},      \\
\label{eq:G5}
\frac{\partial \Psi(E_{ij})}{\partial b_{fj}} &=& -\psi( E_{ij}) a_{if}.     
\label{eq:G6}
\end{eqnarray}

The target function (\ref{eq:G1}) that needs to be minimized 
includes a total of $q(p + n)$ regularization parameters.
The algorithm loops through all the differences $E_{ij}$, minimizing them
as a function of the elements of the two factor matrices $\bA$ and
$\bB$.  If the optimization were to be performed by fixing on $\bB$ and solving the 
optimization with respect to $\bA$ and then reversing the roles of the variables
with the intention to iterate until convergence, there can be difficulties
with convergence given that the two factor matrices are completely unconstrained.
We circumvent this problem by iterating on only some of the elements of $\bA$ before
iterating on some of the elements of $\bB$. This partial updating of $\bA$ 
before a switch to a partial updating of $\bB$ is very effective and is responsible
for the very fast convergence of the process.

This procedure of role reversal between the elements of the two factors matrices
after only partial updating of their elements has been used effectively in 
the context of a recommender system;
see, for example Paterek (2007) and, more recently, Koren (2009),
who used factorization techniques to predict users' preferences
for movies on the Netflix Prize data set.

\begin{figure*}[t]
\centering\includegraphics[width=16cm]{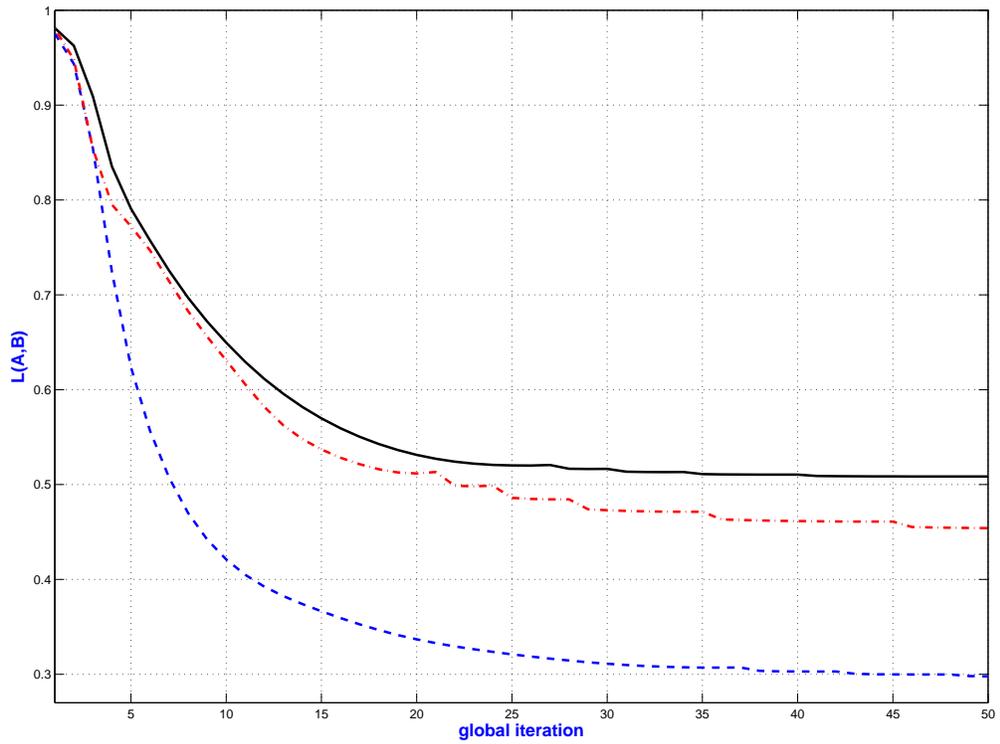}
\caption{Behaviour of the target (\ref{eq:G1}) with squared loss as a function 
of global iteration for
for $q = 10$ metagenes; 
dashed blue, solid black and dot-dashed red lines correspond to the colon, leukaemia and lymphoma cases.} 
\label{fig:target} 
\end{figure*}

It is noted that in the case of the squared loss function, 
we can optimise the value of the step-size. 
However, taking into account the complexity of the model, we recommend
maintaining fixed and small values of the step size or learning rate.
In all our experiments we applied our algorithm
using 100 global iterations with the following regulation parameters.
The initial learning rate $\lambda$ was set at 0.01,
while the correction rate $\xi$ rate was set at 0.75. 
The convergence of the algorithm is illustrated in 
Figure~\ref{fig:target}
for GMF applied to the $2000 \times 62$ data matrix $\bX$ for three data sets,
including the colon cancer data set of Alon et al.\ (1999) with $n=62$ tissues
and $p=2,000$ genes.
The other cancer data sets (leukaemia and lymphoma) are to be described in the next
section.
As pointed out in the introductory section,
our algorithm takes only between 10 and 15 seconds
in performing 300 global iterations
to provide a $q=11$ rank factorization of this data matrix
compared to around 25 minutes to perform 20 global iterations with
non-negative matrix factorization.
We used a Linux computer with speed 3.2GHz, RAM 16GB with the algorithm
using special code written in C).

\section{Application of GMF in Supervised Classification}

In the sequel, we focus on the performance of GMF in its application to some data sets
in the context of supervised classification (discriminant analysis).
In this latter context,
we have an obvious criterion to guide in the choice of the number $q$ of 
metavariables, namely the estimated error rate of the classifier.

Concerning suitable estimates for the error rate of a classifier,
we introduce the following notation.
It is assumed that the observed data points $\bx_1,\,\ldots,\,\bx_n$
come from $g$ possible classes, $C_1,\,\ldots,\,C_g$, with known class labels
specified by $\bz$, where
$$\bz=(\bz_1,\,\ldots,\,\bz_n)^T,$$
and where $\bz_j$ is a $g$-dimensional vector of zeros or ones with its $i$th
element, $z_{ij}$,  defined to be one if $\bx_j$ comes from class $C_i$,
and zero otherwise\, $(i=1,\,\ldots,\,g; j=1,\,\ldots,\,n)$.
For the allocation of an observation $\bx_o$ to one of the $g$ possible
classes, we let
$r(\bx_o; \bX, \bz)$ be a classifier formed from the training data $\bX$
with its known class labels in $\bz$, where $r(\bx_o;\,\bX,\bz)$ equal
to $i$ implies that $\bx_o$ is assigned to class $C_i\,(i=1,\,\ldots,\,g)$.
We shall henceforth abbreviate $r(\bx_o; \bX, \bz)$ to $r(\bx_o; \bX)$
Also, we let
$e(\bX)$ denote an estimate of the error rate of
$r(\bx_o;\bX,\bz)$, 
where dependency of this estimate on $\bz$ is also suppressed for brevity
of expression.
If we use, for example, $n$-fold cross-validation (that is, the
leave-one-out estimate), then
\begin{equation}
e(\bX)= n^{-1}\sum_{i=1}^g\sum_{j=1}^n z_{ij} 
H[i, r(\bx_j;\bX_{(j)})],
\label{eq:6}
\end{equation}
where the function $H[u,v]$ is defined to be equal to 1 
if $u \neq v$, and zero otherwise and where $\bX_{(j)}$ denotes $\bX$ with
$\bx_j$ deleted.
Finally,  
we let $\hbB^{(q)}(\bX)$ denote the solution for $\bB$ when GMF is applied to
$\bX$ for a specified value of $q$.

In the case where the full data matrix $\bX$ is replaced by the reduced
matrix $\hbB^{(q)}(\bX)$ computed for a specified $q$, we can use (\ref{eq:6})
to estimate the expected error rate of the classifier
formed from this reduced set.
An estimate is given by
\begin{equation}
e_1(\hbB^{(q)}(\bX))= n^{-1}\sum_{i=1}^g\sum_{j=1}^n z_{ij}
H[i, r(\bb_j;\hbB_{(j)}^{(q)}(\bX))],
\label{eq:7}
\end{equation}
where $\bb_j$ is the $j$th column of $\hbB^{(q)}(\bX)$ and
$\hbB_{(j)}^{(q)}(\bX)$ denotes $\hbB^{(q)}(\bX)$ with its $j$th column
$\bb_j$ deleted.

As pointed out by Ambroise and McLachlan (2002), this estimate will
provide
an optimistic assessment of the true error rate of the classifier, since
the reduced data matrix $\hbB^{(q)}(\bX)$ should be recomputed on each
fold
of the cross-validation; that is, in the right-hand side of (\ref{eq:7}),
$\hbB^{(q)}_{(j)}(\bX)$ should be replaced by $\hbB^{(q)}(\bX_{(j)})$, 
the reduced data matrix obtained by applying the GMF algortihm to $\bX_{(j)}$,
the data matrix $\bX$ with its $j$th column deleted.
This estimate can be written as 
\begin{equation}
e_2(\hbB^{(q)}(\bX))= n^{-1}\sum_{i=1}^g\sum_{j=1}^n z_{ij}
H[i, r(\bb_j;\hbB^{(q)}(\bX_{(j)}))],
\label{eq:9}
\end{equation}

In order to calculate this estimated error rate
with $n$-fold cross-validation,  it means that
the GMF algorithm has to be run $n$ times in addition to its replication
to the full data set. 
This is feasible given the speed with which the algorithm carries out the GMF.
It should be pointed out that since the GMF does not make use of the known
class labels, the selection bias of the classifier based on the
selected subset of metavariables $\hbB^{(q)}$ will not be nearly as great
in magnitude as with selection methods that use the class labels.
Also, in practice, $n$-fold cross validation can produce an estimate with
too much variability and so five- or ten-fold cross validation is often used
in a variance versus bias tradeoff (Ambroise and McLachlan, 2002).

We can choose the final value of $q$ by taking it to be the value
$q_o$ that minimizes the estimated error rate $e_2(\hbB^{(q)}(\bX))$;
\begin{equation}
q_o=\arg \min_{q \in Q} e_2(\hbB^{(q)}(\bX_{(j)}),
\label{eq:8}
\end{equation}
where $Q$ denotes the set of values considered for $q$.
However, there is still a selection bias if we use
\begin{equation}
e_2(\hbB^{(q_o)}(\bX))= n^{-1}\sum_{i=1}^g\sum_{j=1}^n z_{ij}
H[i, r(\bb_j;\hbB^{(q_o)}(\bX_{(j)}))],
\label{eq:9a}
\end{equation}
to estimate the error rate of the classifier based on the reduced set
with the smallest error rate over the values of $q$ considered;
see, for example, (Wood et al., 2007; Zhu et al., 2008).
We can correct for this bias by using the estimate
\begin{equation}
e_3(\hbB^{(q_o)})= n^{-1}\sum_{i=1}^g\sum_{j=1}^n z_{ij}
H[i, r(\bb_j;\hbB^{(q_{oj})}(\bX_{(j)}))],
\label{eq:10}
\end{equation}
where
\begin{equation}
q_{oj}= \arg \min_{q \in Q} \sum_{i=1}^g 
\sum_{\begin{subarray}{l}j'=1\\j' \neq j\end{subarray}}^{n}
\frac{z_{ij'}H[i,r(\bb_{j'};\,\bB^{(q)}(\bX_{(j,j')}))]}{n-1},
\label{eq:11}
\end{equation}
and 
$\bX_{(j,j')}$ denotes the data matrix $\bX$ with $\bx_j$ and $\bx_{j'}$
deleted.

It can be seen from (\ref{eq:11}) that in order to calculate the
cross-validated estimate (\ref{eq:10}), we need to perform the GMF
$n(n-1)$ times in addition to the original application to the full data
set $\bX$.
This is still feasible since GMF can be implemented so quickly, although
the total computational time becomes large as $n$ increases.
As noted above, using, say, ten-fold cross-validation would reduce
the number of times that
GMF has to be employed.
In the data sets considered here, the increase in the estimated error rate
given by  the use of $e_3(\hbB^{(q_o)})$ over 
(\ref{eq:9a}) was very small (not of practical significance).

\section{Supervised Classification of Some Cancer Data Sets}

We shall demonstrate the application of the GMF for dimension reduction in the
context of supervised classification
of five cancer data sets that have been
commonly analysed in the bioinformatics literature,
as briefly described in the following section.

\subsection{Five Data Sets}

For the colon data set (Alon et al., 1999)
the data matrix $\bX$
contains the expression levels of $p$=2000 genes in each of $n$=62 tissue samples 
consisting of $n_1$=40 tumours and $n_2$=22 normals.

The data matrix for the leukaemia data set (Golub et al., 1999)
contains the expression levels 
of $p$ = 7129 genes for each of $n$ = 72 patients, consisting of $n_1$=47 patients 
suffering from acute lymphoblastic leukaemia (ALL) and 
$n_2$= 25 patients suffering
from acute myeloid leukaemia (AML).

We followed the pre-processing steps of (Golub et al., 1999) applied to the leukaemia set:
1) thresholding: floor of 1 and ceiling of 20000; 
2) filtering: exclusion of genes with max/min $\leq$ 2 and 
(max - min) $\leq$ 100, where max and min refer respectively to the maximum and minimum 
expression levels of a particular gene across the tissue samples.
This left us with $p = 1896$ genes.
In addition, the natural logarithm of the expression levels was taken. 

The data matrix for the lymphoma data set (Alizadeh et al., 2000)
contains the gene expression levels of the three most prevalent adult 
lymphoid malignancies: $n_1$= 42 samples of diffuse large B-cell lymphoma (DLCL), 
$n_2$= 9 samples of follicular lymphoma (FL), and $n_3$= 11 
samples of chronic lymphocytic leukaemia (CLL).
The total sample size is thus $n$= 62 and there are $p$ = 4026 genes.

The Sharma data set was described in (Sharma et al., 2005) and contains the expression levels
(mRNA) of $p$= 1368 genes for each of 60 blood samples taken from 56 women.
Each sample was labelled by clinicians, with $n_1$=24 labelled as having breast 
cancer and $n_2$ = 36 labelled as not having it.
Some of the samples were analysed more than once in separate batches giving a
total of $n$= 102 labelled samples.

The fifth data set (Khan et al., 2001) contains the expression levels of
$p$= 2308 genes for each of $n$ =  83 tissue samples,
each from a child who was determined by clinicians
to have a type of small round blue cell tumour.
This includes the following $g$=4 classes:
neuroblastoma (N), rhabdomyosarcoma (R), Burkitt lymphoma (B) and the Ewing sarcoma (E).
The numbers in each class are:  N($n_1$=18), R($n_2$=25), B($n_3$ = 11), and
E($n_4$=29).

We applied double normalization to each data set. 
Firstly, we normalized each column 
to have means zero and unit standard deviations. 
Then we applied the same normalization to each row.

\subsection{Error rates for classifiers formed on the basis of metagenes}

\begin{figure*}[t]
\centering\includegraphics[width=16cm]{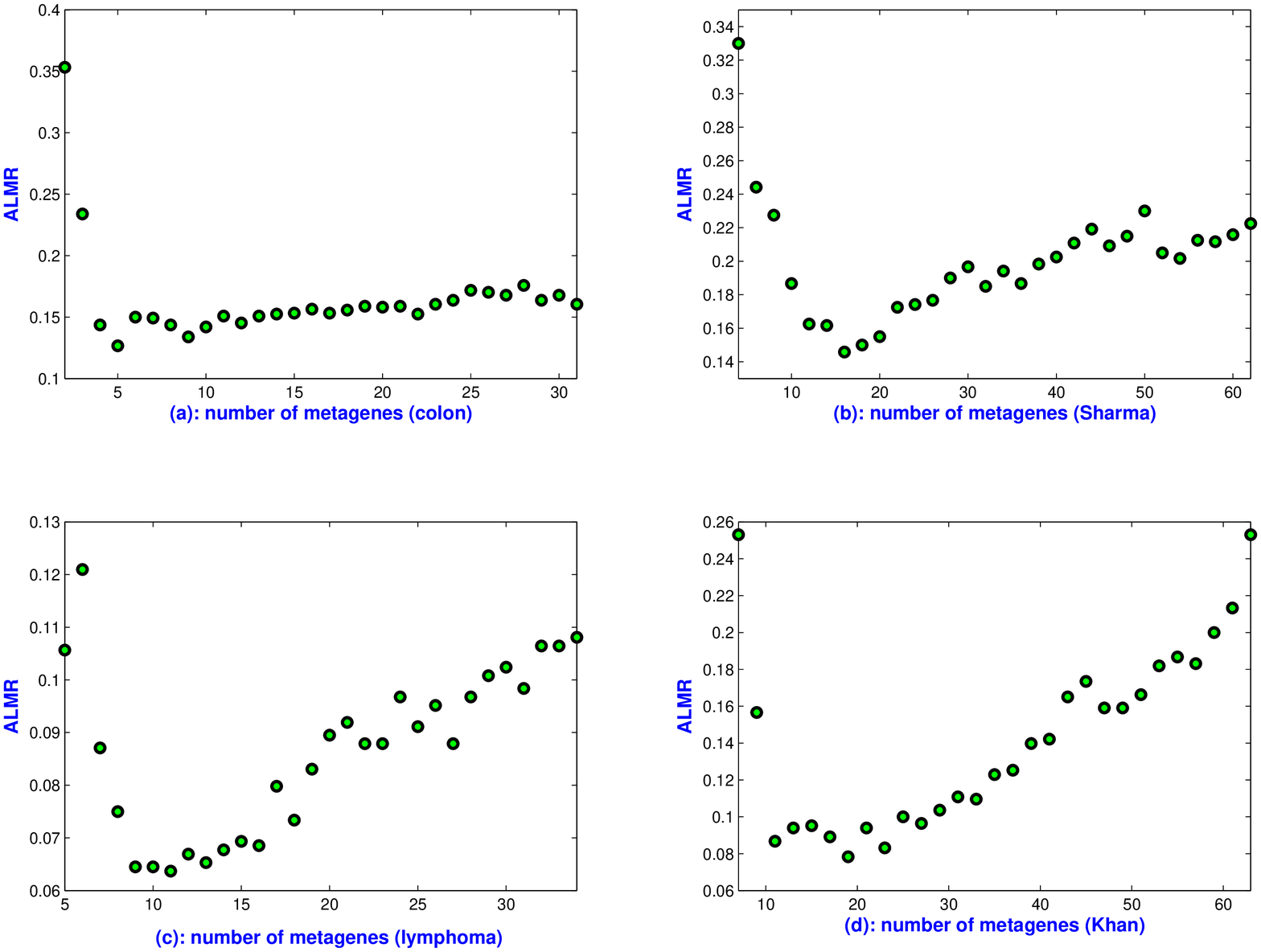}
\caption{The estimated errror rate $e_1$ as a function of 
the number $q$ of metagenes.}
\label{fig:almr}
\end{figure*}

In Figure~\ref{fig:almr}, we plot the cross-validated error rate $e_1$ versus
the number of metagenes $q$ for four of the five data sets, 
using the support vector machine (SVM) 
in the case of $g=2$ classes and (multinomial) logistic regression (LR) in the
case of $g>2$ classes.  We did not plot $e_1$ for the leukaemia data set
as it was close to zero if $q \geq 10$.

In Table~\ref{tab:1}, 
we list the cross-validated error rates $e_1$ and its bias-corrected version
$e_2$ for each of the four data sets, where the classifier (SVM or MLR) 
is formed on the basis of $q$ metagenes.
To give some guide as to the level of performance of the performance of these
classifiers, we also list the value of the error rate using the 
nearest-shrunken centroids method (Tibshirani et al., 2002).
The bias-corrected error rate $e_2$ is smaller than that of the NSC method for all 
but one of the data sets (the lymphoma set).
The estimated error rate for the nearest-shrunken method corresponds to 
$e_2$ in that it can be regarded as an almost unbiased estimate for a given 
subset of the genes, but it has not been corrected for bias over the set 
of values $q$ considered; see Wood et al.\ (2007) and Zhu et al.\ (2008).

\begin{figure*}[t]
\centering\includegraphics[scale=0.45]{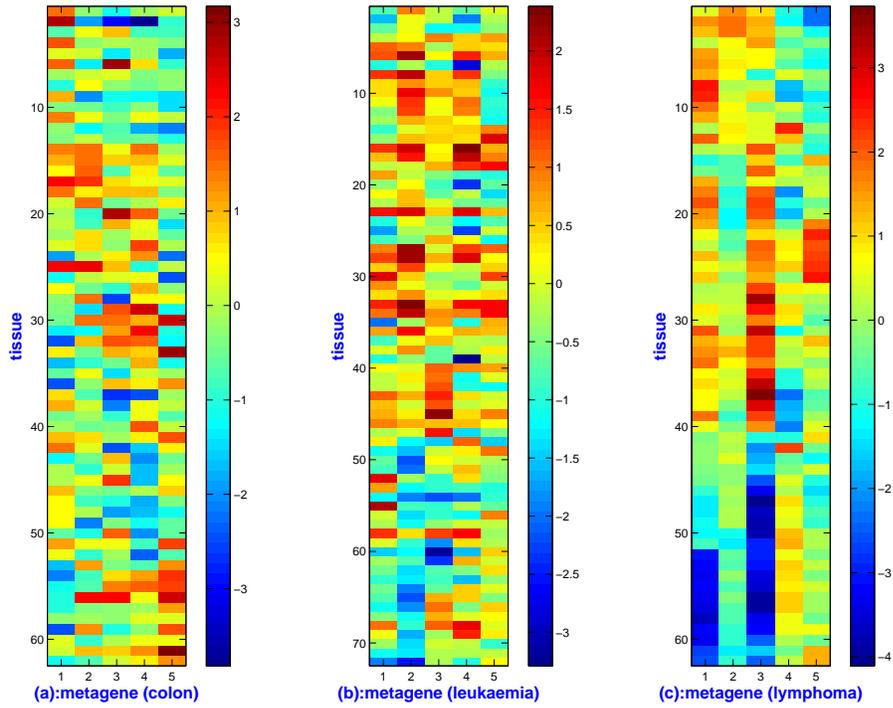}
\caption{Images of the matrix $\bB$ for
$q = 5$: (a) colon (sorted from the top: 40 positive then 22 negative), 
(b) leukaemia (sorted from the top: 47 ALL, then 25 AML) and 
(c) lymphoma (sorted from the top: 42 DLCL, then 9 FL, last 11 CLL). 
All three matrices were produced using the GMF algorithm with 100 global 
iterations as described in the text.}
\label{fig:metagenes3}
\end{figure*}

On the question of which metavariables (metagenes) are useful in the
discriminatory process, an inspection of the heat maps (coloured
lots of each metagene value for each tissue) 
can be useful, but not always. 
To illustrate this, we have plotted the heat maps in
Figure~\ref{fig:metagenes3} for three of the data sets.
In this figure, we have sorted the tissues into their classes in order
to consider visual differences between the patterns.
In the case of the colon data in
Figure~\ref{fig:metagenes3}(a), we cannot see clear separation
of the negative and positive classes.
In contrast, in the case of the leukaemia data in 
Figure~\ref{fig:metagenes3}(b), metagene N2
separates the first 47 tissues (from the top) from
the remaining 25 tissues with only one exception.
It is tissue 58, which is the only one misclassified tissue
in Table~\ref{tab:1} (cases $q = 3, 4$).
Similarly, in the case of the lymphoma data in
Figure~\ref{fig:metagenes3}(c),
metagene N1
separates clearly CLL from the remaining two classes.
Further, metagene N3 separates DLCL from
the remaining two classes.


To assist with the interpretation of the metagenes, we can examine the 
Gene Ontology (GO) (The GO Consortium, 2009) 
and the pathway records of the Kyoto Encyclopedia of Genes and
Genomes (KEGG) (Kanehisa et al., 2010) for those genes that have high (absolute) correlations with 
the metagenes.

\section{Conclusions}

We have presented an algorithm for performing extremely fast general matrix
factorization (GMF) of a high-dimensional data matrix $\bX$.
In practice some form of dimension reduction is invariably 
needed if standard or even
nonstandard methods of statistical analysis are to be employed to gain meaningful
insight from high-dimensional data matrices.
The algorithm undertakes the factorization using gradient-based optimization for
an arbitrary (differentiable) loss function.
The  $p \times n$ data matrix  $\bX$ is approximated by the product of two matrices, 
$\hbA\hbB$, where the $q \times n$ factor matrix $\hbB$ can be used in place of 
$\bX$ for a specified value of $q$ taken to be much smaller than $p$.
The $n$ columns of $\hbB$ contain the values of the $q$ metavariables for each of
the $n$ observed data vectors.
The stability of the algorithm depends essentially on a properly selected
learning rate, which must not be too big.
We can provide additional functions so that the learning rate will be reduced
or increased depending on the current performance.

To demonstrate the usefulness of the reduced data matrix $\hbB$ 
in the context of supervised classification, we applied it to the data matrices from
five cancer data matrices of microarray gene expressions that have been commonly 
analysed in the medical and scientific literature.
The  classification of the microarrays (tissue samples) containing these
gene expressions are of known classification with respect to $g$ classes, where
$g$ varies from 2 to 4.
The results suggest that GMF as implemented by our algorithm is effective 
in providing a reduced data matrix for their subsequent use in forming 
a classifier that was taken to be either the SVM in the case of $g=2$ classes 
or logistic regression for $g>2$ classes.

Some main issues associated with the use of GMF are the choice of the number $q$
of metavariables (latent factors),  the interpretability of the metavariables, 
and the need for prefiltering of the variables before the factorization. 
These issues are beyond
the scope of this paper.

But briefly on these issues here, the choice of the number of metavariables $q$  
in supervised classification can be based on the estimated error rate of the
classifier formed from the specified number of metavariables.
The situation is not as straightforward in the context of cluster analysis where
there are no training data of known origin even if there were some {\it a priori}
knowledge about the existence of some group structure in the data. 
One way to
proceed in this context is to use the stability in the clustering as the level of
$q$ is varied as a guide to its final choice (Brunet et al., 2004, 
Tamayo et al., 2007). 
The question of whether there is a need for prefiltering 
in the context of cluster analysis 
has been considered recently by Zheng et al.\ (2009).

The problem of interpretability of the metavariables is generally not as 
straightforward as with NMF's since the latter are non-subtractive combinations
of the original variables (Zheng et al., 2009).
In general, we can calculate the correlations between the original variables and
the metavariables. 
For microarray data, we are currently designing a program that automatically
attempts to establish links between genes highly correlated with a metagene
and the Gene Ontology (The GO Consortium, 2009) and the the pathway records of the Kyoto
Encyclopedia of Genes and Genomes (Kanehisa et al., 2010).

\section*{References}
\noindent
Lee, D.D.,   Seung, H.S., 1999. Learning the parts of objects by
non-negative matrix factorization.
Nature 401, 788--791.\\

\noindent
Golub, G.,   van Loan, C., 1983. Matrix Computations.
Johns Hopkins University Press, Baltimore.\\

\noindent
Eckart, C.,   Young, G., 1936.
The approximation of one matrix by another of low rank.
Psychometrika  1, 211--218.\\

\noindent
Alon, U., Barkai, N.,
Notterman, D.A., Gish, K.,
Ybarra, S., Mack, D.,
Levine, A.J., 1999.
Broad patterns of gene expression revealed by clustering analysis of
tumor and normal colon tissues probed by oligonucleotide arrays.
Proceedings of the National Academy of Sciences USA  96, 6745--6750.\\

\noindent
Lee, D.D.,   Seung, H.S., 2001. Algorithms for non-negative matrix
factorization. Advances in Neural Information Processing System  13, MIT Press.\\

\noindent
Li, S.Z., Hou, X.W., Zhang, H.J.,   Cheng, Q.S., 2001. Learning
spatially localized parts-based representation.
In Proceedings of IEEE International Conference on Computer
Vision and Pattern Recognition Volume I, Hawaii,  pp.207--212.\\

\noindent 
Donoho, D.,   Stodden, V., 2004. When does non-negative matrix
factorization give a correct decomposition into parts ?
In   Advances in Neural Information Processing Systems 16,
Cambridge, MA: MIT Press. \\

\noindent
Gao, Y.,   Church, G., 2005. Improving molecular cancer class discovery 
through sparse non-negative matrix factorization.
Bioinformatics 21, 3970--3975.\\

\noindent 
Fogel, P., Young, S.S., Hawkins, D.M.,   Ledirac, N., 2007.
Inferential, robust non-negative matrix factorization analysis of microarray data.
Bioinformatics  23, 44--49.\\

\noindent
Cichocki, A., Zdunek, R.,  Phan, A.H.,    Amari, S-I., 2010.
Nonnegative Matrix and Tensor Factorizations. 
Wiley, Chichester\\

\noindent 
Ding, C., Li, T.,   Jordan, M.I., 2010. Convex and semi-nonnegative
matrix factorizations.  IEEE Transactions on Pattern Analysis 
and Machine Intelligence  32, 45--55.\\

\noindent 
Witten, D.M., Tibshirani, R.,   Hastie, R., 2009. 
A penalized matrix decomposition, with applications to sparse principal
components and canonical correlation analysis.   Biostatistics  10, 515--534.\\

\noindent
Nikulin, V.,    McLachlan, G.J., 2009.
On a general method for matrix factorization applied to supervised
classification.
In   Proceedings of 2009 IEEE International Conference on
Bioinformatics and Biomedicine Workshop, Washington, D.C. ,
J Chen et al. (Eds.).
Los Alamitos, California: IEEE, pp.\ 43--48.\\

\noindent
Paterek, A., 2007.
Improving regularized singular value decomposition for collaborative filtering.
  KDD Cup , San Jose, CA: ACM, pp.\ 39--42.\\

\noindent
Koren, Y., 2009.
Collaborative filtering with temporal dynamics. In
  Proceedings of the 15th ACM SIGKDD Conference on Knowledge Discovery
and Data Mining , Paris, pp.\ 447--455.\\

\noindent
Ambroise, C.,   McLachlan G.J., 2002.
Selection bias in gene expression on the basis of microarray gene expression data.
  Proceedings of the National Academy of Sciences USA  99, 6562--6566.\\

\noindent
Wood, I., Visscher, P.,   Mengersen, K., 2007. Classification based
upon expression data: bias and precision of error rates.
  Bioinformatics  23, 1363--1370.\\

\noindent
Zhu, J.X., McLachlan, G.J., Ben-Tovim, L., Wood, I., 2008.
On selection biases with prediction rules formed from gene expression
data. Journal of Statistical Planning and Inference  38, 374--386.\\

\noindent
Golub, T.R., Slonim D.K.,
Tamayo, P., Huard, C.,
Gassenbeck, M.,
Mesirov, J.P., Coller, H.,
Loh, M.L., Downing, J.R.,
Caligiuri, M.A., et al. 1999.
Molecular classification of cancer: class discovery.
Science 286, 531--537.\\

\noindent
Dudoit, S., Fridlyand, J.,   Speed, T.P., 2002. Comparison of
discrimination methods for the classification of tumors using gene
expression data. Journal of the American Statistical Association  
97, 77--87.\\

\noindent
Alizadeh, A., Eisen, M.B.,
Davis, R.E., Ma, C.,
Lossos, I.S., Rosenwal. A.,
Boldrick, J.C., Sabet, H.,
Tran, T., Yu, X., et al. 2000.
Distinct types of diffuse large B-cell lymphoma identified by gene
expression profiling.
Nature  403, 503--511.\\

\noindent
Sharma, P., Tibshirani, R., Skaane, P., Urdal, P.,   Berghagen, H., 2005.
Early detection of breast cancer based on gene-expression patterns in peripheral
blood cells.
Breast Cancer Research  7, R634--R644.\\

\noindent
Khan, J., Wei, J., Ringner, M., Saal, L., Ladanyi, M., Westermann, F., Berthold, F., 
Schwab, M., 2001.
Classification and diagnostic prediction of cancers using gene expression
profiling and artificial neural networks.
Nature Medicine  7, 673--679.\\

\noindent
Tibshirani, R., Hastie, T., Narasimhan, B.,   Chu, G., 2002.
Diagnosis of multiple cancer types by shrunken centroids of gene expression.
Proceedings of the National Academy of Sciences USA  99, 6567--6572.\\

\noindent
The GO Consortium. 2009. Gene ontolog  [Online]. Available:\newline
\url{http://www.geneontology.org}\\

\noindent
Kanehisa, M., Goto, S., Furumichi, M., Tanabe, M., Hirakawa, M., 2010.
KEGG for representation and analysis of molecular networks 
involving diseases and drugs.
Nucleic Acids Research 38, D355--D360.\\

\noindent
Brunet, J-P., Tamayo, P., Golub, T.R., Mesirov, J.P., 2004.
Metagene projection for cross-platform, cross-species characterization of global
transcriptional states.
Proceedings of the National Academy of Sciences USA  101, 4164--4169.\\

\noindent
Tamayo, P., Brunet, J-P., Scanfeld, D., Ebert, B.L., Gillette, M.A., Roberts, C.W.M.,   Mesirov, J.P., 2007.
Proceedings of the National Academy of Sciences USA  104, 5959--5964.\\

\noindent
Zheng. C., Huang, D., Zhang, L.,   Kong, X., 2009. 
Tumor clustering using nonnegative matrix factorizations with gene selection.
IEEE Transactions on Information Technology in Biomedicine  13, 599-607.\\

\begin{table*}
\def~{\hphantom{0}}
\caption{Some selected experimental results,
where numbers in brackets in the first column indicate numbers of classes in 
the corresponding data set,
and numbers of misclassified entries in the fourth, fifth and sixth columns.
Results in the sixth column \lq\lq NSC" were obtained using nearest-shrunken 
centroids method
with threshold parameter $\Delta$ as it was described in Tibshirani et
al., (2002).
The column $p_s$ indicates the number of used/selected features.
}{%

\begin{tabular}{lcccc|ccc} \\
Data & Model &  q & $e_1$ & $e_2$ & NSC & $p_s$ & $\Delta$ \\[10pt]
\hline
Colon (2) & SVM & 8 & 0.0806 (5) & 0.1129 (7) & 0.129 (8) & 141 & 1.3 \\
Leukaemia (2) &  SVM &  25 & 0 (0) & 0.0139 (1) & 0.0139 (1) & 73 & 1.9 \\
Lymphoma (3) & MLR & 10 & 0.0322 (2) & 0.0322 (2) & 0.0161 (1) & 3336 & 0.8 \\
Breast (2) & SVM & 18 & 0.1 (6) & 0.15 (9) & 0.2 (12) & 53 & 1.2 \\
Blue cell tumour (4) & MLR & 21 & 0.0241 (2) & 0.0482 (4) & 0.0602 (5) & 464 & 1.8
\label{tab:1}
\end{tabular}}
\end{table*}

\end{document}